\documentclass[
aps,%
12pt,%
final,%
notitlepage,%
oneside,%
onecolumn,%
nobibnotes,%
nofootinbib,%
superscriptaddress,%
noshowpacs,%
centertags]%
{revtex4}

\def\la{\mathrel{\mathpalette\fun <}}

\def\fun#1#2{\lower3.6pt\vbox{\baselineskip0pt\lineskip.9pt
\ialign{$\mathsurround=0pt#1\hfil##\hfil$\crcr#2\crcr\sim\crcr}}}

\begin{document}

\title{   Mass Spectrum and Number of Light Neutrinos: An Attempt
                  of the Gauge Explanation}

\author{\firstname{\bf I. T. Dyatlov} }

\affiliation{Petersburg Nuclear Physics Institute, RAS, Gatchina 188350,
                           Russia}

\begin{abstract}\noindent
Symplectic flavour symmetry group $Sp(n/2)$ ($n$  is  even)
     of  $n$  Majorana  states does not  allow  for  invariant
     Majorana  masses.  Only  specific  mass  matrices  with
     diagonal and nondiagonal elements are possible here. As
     a  result  of the spontaneous violation of flavour  and
     chiral symmetries, a mass matrix could appear only  for
     the number of flavours $n = 6$ and only together with $R,L-$
     symmetry  violation (i.e., parity violation). The  see-
     saw mechanism produces here three light and three heavy
     Dirac  particles  (neutrinos). The peculiarity  of  the
     observed  light neutrino spectrum – two states  located
     far  from  the third one – can be explained by  certain
     simple  properties of mass matrices appearing in $Sp(3)$.
     The  ordering of the states corresponds to normal  mass
     hierarchy. 
Situation, when neutrino mass differences are
significantly less than masses themselves, appears to be unrealizable
here. 
Mixing  angles for  neutrinos  can  not  be determined  without
     understanding formation  mechanisms for  charged lepton spectrum
     and  Majorana  state  weak currents.

\end{abstract}

\maketitle

\section{Introduction}

      Neutrinos  have  a mass. The spectrum  and  mixing  of
neutrinos  are based on other principles than the respective
properties  of  charged, and therefore  compulsorily  Dirac,
massive  states \cite{1}.  That is why,  the  thought  that  the
Majorana properties (which only neutrinos may have)  play  a
crucial  part  in  the  origin  and  character  of  neutrino
spectrum is gaining wider support \cite{2}.

      If  so, the mechanism of neutrino mass generation  has
little  in  common with the mechanism of charged Dirac  mass
generation.  Mixing angles are the simultaneous  consequence
of  both  mechanisms and therefore can not provide neutrino-
specific information on the mass generation model. There are
only  the smallness of the neutrino mass as opposed  to  any
other particles, the number of light neutrinos  $N_\nu = 3$,  and
the  large ratio of mass squared differences for these light
neutrinos   (1)   that   indeed   can   be   considered   as
characteristics of the neutrino mechanism .  Using  standard
conventions \cite{1}, one has
\begin{equation}
23\la \frac{\Delta m_{23}^2}{\Delta m_{12}^2}\la 40
\end{equation}
Relation  (1)  indicates that one of  the  states,  "3",  is
located rather far from the other two, which are very  close
to  each  other. One distinguishes here direct and  inverted
hierarchies \cite{2,3}.

      The smallness of neutrino masses is the only aspect to
this phenomenon that has a fine consistent explanation.  The
see-saw  mechanism,  developed  and  investigated  by   many
authors  (see  review \cite{3}),  considers  the  smallness   of
neutrino masses as a result of the existence of a very  high
energy  scale. All proposed models appear with participation
of scalar interactions and scalar (Higgs) particles; they do
not  explain  relation (1), nor do they distinguish  between
the types of hierarchy.

       At the same time, all observed interactions are based
on  the  exchange of gauge bosons (with a vector  or  tensor
spin).  The  question  is whether mass  formation  phenomena
could   similarly  be  attributed   only  to   local   gauge
mechanisms. This, of course, would imply a dynamical way for
mass  formation,  and result in inaccessible nonperturbative
solutions. What only can be achieved really by this approach
are  direct symmetric consequences, and they are the subject
of this paper.

       Let  us  consider $n$ Majorana states. For  the  flavor
nonabelian    gauge  group  of  symplectic   transformations
$Sp(n/2)$,  $n=2,4,6,\ldots$,  the invariant Majorana masses,  both
for  chiral  right R  and left L particles, are  identically
equal to zero. Thus, it is only the whole mass matrix,  with
both  diagonal  and  nondiagonal  elements,  that  could  be
produced  here  by  dynamical spontaneous breaking.  On  the
other  hand,  the  Dirac part of the complete  $(R,L)$  matrix
could be invariant.

The principal accomplishment of the present approach is
equations (26)-(28) in Section 5. They provide a  choice
of  conditions under which the spontaneous appearance of the
mass  matrix  becomes  possible in $Sp(n/2)$.  There  are  two
possible solutions: one is $R,L-$symmetry at $n = 2$, $Sp(1)$, and
the  other, the spontaneous breaking of $R,L-$symmetry at $n=6$,
$Sp(3)$.  All neutrinos  necessarily appear as Dirac ones.  In
the  physical  sense, the second solution, $n=6$,  is  more
interesting. It creates conditions under which  the  see-saw
mechanism  divides the six Dirac neutrinos into three  light
and three heavy ones. Note that these conditions can only be
realized  for  Majorana states: only then     equations  for
spontaneous  mass matrices become self-consistent,  although
their  solutions  resolve  into  exclusively  Dirac  massive
particles.

Majorana mass distributions in $Sp(3)$ and the action of
the  see-saw mechanism permit such a disposition  for  light
particles where one state is located far from the other two.
Such   a  spectrum  can  be  explained  by  a  rather  usual
distribution  of  the  roots  of  the  cubic  characteristic
equation  for Majorana mass squared which results in  $Sp(3)$.
The quantities of the system do not require any fine tuning,
besides  providing the see-saw situation. The light neutrino
mass hierarchy is normal.

In Section 2, we discuss reasons for selecting $Sp(n/2)$
as  a  flavour symmetry in the Majorana problem.  Section  3
investigates   properties  of  mass   matrices   which   are
acceptable  in  $Sp(n/2)$.  Section 4 proposes a  gauge  model
considered  for  a hypothesis of spontaneous  generation  of
mass  matrices. It describes the properties that would allow
dynamical   violation  of  flavour  and  chiral  symmetries.
Section  5 discusses the conditions under which the proposed
mechanism  would  work. Dirac states  corresponding  to  the
Majorana  spectrum  $Sp(3)$  are  constructed  in  Section  6.
Section  7 considers possible explanations in $Sp(3)$ for  the
observed  light neutrino spectrum . Section 8,  Conclusions,
describes  difficulties in transition to a realistic  model,
which includes charged leptons, i.e., the whole set of  weak
processes.

\section{ Choosing Gauge Group}

      Majorana  state  flavours cannot transform  under  any
representation  of  their symmetry  group  $G_F$.  Indeed,  the
identity  of  the particle and antiparticle,  which  relates
spinors $\Psi_a(x)$ and $C\bar \Psi^{Ta}(x)$ (where $a$ is a flavour
index, $a=1,2,\ldots,n$),  requires that conjugate (contragredient
\cite{4}) representations be equivalent.  Conjugate   representations
are related   by   root reflections  at  the coordinate origin  and by
changes  in infinitesimal  operators: $t^A\rightarrow
  -t^{AT} (t^{A^+}=t^A)$ \cite{4}.
  For equivalent conjugate representations, there exists a
matrix $n\times n$, $h$ , that allows obtaining $-t^{AT}$ from
$t^A$:
\begin{equation} h^+t^Ah=-t^{AT}\,.
\end{equation}
This  matrix raises and lowers $\Psi$ indices, and its properties
and notations are as follows:
\begin{equation} h=\{h_{ab}\}\,,\quad h^+=\{h^{ab}\}\,,\quad
hh^+=1\,,\quad h^{ab}=\pm h_{ab}\,,\quad h^T=\pm h\,.  \end{equation}
The  designation  of  $h$ - is  to renumber  roots.  The  right
diagonal of the matrix $h$ contains exclusively $\pm 1$ elements.

Let  us first consider $n$ massless Majorana states.  In
terms  of  common chiral operators $\psi_{RL}=1/2(1\pm\gamma_5)\psi$
,  there
are  two operators covariant with respect to $G_F$ group  which
can  be  called  "Majorana-like". In a four-component  form,
they are:
\begin{equation}
\Psi_{(R,L)a}(x)=\psi_{(R,L)a}(x)+(1,\gamma_5)h_{ab}C\bar
\psi_{(R,L)}^{Tb}(x)\,.
\end{equation}

The factor $(1,\gamma_5)$ is associated with the sign of
$h^T = \pm h$ in
Formula (3): at $h^T = h$ one should take the unity matrix, and
at  $h^T=-h$, the matrix $\gamma_5$. Normalization in (4) is chosen
so  that the energy of free massless Majorana particles  can
be expressed in a usual way for neutral states:

\begin{equation}
\bar \psi^a\hat p\psi_a=\frac 12 \bar \Psi^a\hat p\Psi_a\,.
\end{equation}

      States (4) transform under the same representation  of
the   flavour   group  $G_F$  as $\psi_{(R,L)a}(x)$   and  demonstrate
"Majorana" properties:
\begin{equation}
\Psi_{(R,L)}(x)=(1,\gamma_5)hC\bar \Psi_{(R,L)}^T(x)\,,
\end{equation}
$$C^+=C^T=-C\,.$$

In  Eq.  (6),  one  observes  a complete  analogy  with  the
extended charge parity ($G-$parity) \cite{6}: the transition to the
antiparticle  takes  place  simultaneously  with  the  group
operation.  The appearance of $\gamma_5$ at $h^T = -h$ has significance
only  for  massive states $(m\rightarrow -m)$, and ultimately leads  to
important physical consequences (see sections 3,6,7).

Complex  conjugate representations are equivalent  in
symmetric representations (with root reflection). These  are
primarily adjoint representations of all groups. There  also
exist fundamental representations which demonstrate the same
property,   i.e.,   representations  of   orthogonal $O(n)$,
symplectic   $Sp(n/2)$  (using  designations  of \cite{5}),   and
exceptional  groups \cite{4,5}. The $n-$dimensional representation
$Sp(n/2)$, $n = 2, 4, 6,\ldots$, has an antisymmetric $h$:
\begin{equation}
h^T=-h\,, \qquad h^+=-h\,.
\end{equation}
All other cases are symmetrical matrices $h$.

Of  greatest  interest in (7) is that the  dynamical
violation  of flavour and chiral symmetries gives rise  here
to  a whole Majorana mass matrix. Indeed, only a matrix with
non-zero  diagonal and nondiagonal elements  can  result  in
this  case because the invariant Majorana mass here is equal
to zero:
\begin{equation}
\bar \Psi_R^a(x)\Psi_{Ra}(x)=-\bar \psi_R(x)hC\bar \psi_R^T(x)-
\psi_R^T(x)h^+C\psi_R(x)=0\,,
\end{equation}
which is similar also for $\Psi_L$. Identity (8) results from  the
anticommutation  of  operators  $\psi_R$  and  matrix $h$ and $C$
properties. This identity implies that, under condition (7),
the  appearance of Majorana masses is only possible with the
full breaking of flavour symmetry, to the point where a mass
matrix  without residual symmetries is created. This  matrix
will  immediately presented states with various mass  values
in the spectrum.

On  the  other hand the Dirac mass  can   exist  in
$Sp(n/2)$ in invariant form

\begin{equation}
\bar \Psi_R^a(x)\Psi_{La}(x)=\bar \psi_R^a(x)\psi_{La}(x)-\bar
\psi_L^a(x)\psi_{Ra}(x)\neq 0\,\footnote{The form (7) corresponds to
the imaginary Dirac mass. Transition to real mass values occures after
redetermination of
 $\Psi_{(R,L)}$ in Eq.(4): $\Psi\rightarrow(i, \mbox{¨«¨}
\gamma_5)\Psi$ (sect. 3).}.
\end{equation}
Under  symmetric representations $h = h^T$, Majorana and  Dirac
masses  may both be present even without symmetry  breaking.
This would be the simplest way for spontaneous violation  to
occur,  with the result being equal masses for all  flavours
rather than a mass matrix.

We,  therefore,  shall consider those  possibilities
that   may   present,  for  neutrino  masses,  a  hypothetic
spontaneous violation of the flavour symmetry $Sp(n/2)$  under
the fundamental representation $n$. Note that this symmetry of
leptons  can also be the local gauge symmetry of interaction
with  the vector field $F_\mu^A$, $A = 1,2,\ldots, n(n + 1)/2$.
Inclusion of  this interaction does not result in new anomalies,  even
if  weak  interactions are taken into account. Full symmetry
breaking leads to the absence of Goldstone particles and  to
the formation of heavy masses simultaneously for all $F_\mu^A$.
       Let us first consider the properties of an acceptable
mass matrix in the gauge-invariant scheme.

\section{  Majorana Mass Matrix Properties}

        In essence, dynamical flavour symmetry violation  is
the   generation   of  vacuum  averages of $\bar \Psi^a(x)\Psi_b(x)$
combinations for $R-$ and $L-$operators. According  to  (8),  the
trace  of  such matrix will be equal to zero for $RR-$ and $LL-$
systems;  however, individual, or all, matrix elements  will
be other than zero. For $RR-$ and $LL-$systems, we have:
\begin{equation}
(M_{RR})_a\,^b=\left<\bar \Psi_R^b(x)\Psi_{Ra}(x)\right>,\,\,\,\,\,
(M_{LL})_a\,^b=\left<\bar \Psi_L^b(x)\Psi_{La}(x)\right>.
\end{equation}

        Let  us  consider a symmetrical matrix $M_a^b$ which  is
real,  as  equations  for spontaneous  violation  parameters
("gap" equations for mass quantities [7]) will by all  means
be real (with CP conservation).

   Majorana   conditions  (6)  for  the   case   under
consideration are:
\begin{equation}
\Psi_R(x)=\gamma_5hC\bar \Psi_R^T(x)\,,
\end{equation}
For the $L$ operator, we have:
\begin{equation}
\Psi_L(x)=-\gamma_5hC\bar \Psi_L^T(x)\,.
\end{equation}

Form  (12)  depends on the phase selected,  which,  for  the
given  form  (11),  makes  the Dirac  mass  (9)  also  real.
Equations   (11)   and   (12),   and   the   properties   of
anticommutation  $\Psi$  result  in the following,  non-covariant
form of the relation between the matrix elements $M_a^b$ (for $RR$
and $LL$):
\begin{equation} M_a\,^b=-h_{aa'}M_{a'}\,^{b'}h^{b'b}\,.
\end{equation}
The  covariant form of this relation is $M^{+T} =-h^+Mh$.  The
matrix h elements are:
\begin{equation} h_{ab}=-h^{ab}=(-1)^a\delta_{n+1-a,b}\,,\quad
a,b=1,2,\ldots,n\,.  \end{equation}

     Trace of the matrix $M$ evidently vanishes (13). As well,
condition  (13) implies vanishing of sums of  all  principal
minors  of odd order for matrices (10). To prove it,  it  is
sufficient to write the sought sum in the following form:
$$\frac{1}{n_{M}!}\varepsilon^{ab\ldots a_1a_2\ldots}M_{a_1}\,^{b_1}
M_{a_2}\,^{b_2}\ldots \varepsilon_{ab\ldots b_1b_2\ldots}\,,$$
where  the  letters without indices, $ab\ldots$, are rows (columns)
that  are  not  used in the calculation of the minors  under
consideration, and to apply relation (13) and the formula:
$$\varepsilon^{a_1a_2\ldots}h_{a_1a'_1}h_{a_2a'_2}\ldots=
\varepsilon_{a'_1a'_2\ldots}\,.$$

      The  eigenvalue equation for matrix (10) will  contain
only even powers of eigenvalues $M_f$. For each of the matrices
$M_{RR}$  and $M_{LL}$, there are $n/2$ states which differ in the value
of  $M_j^2$.  Thus, since a pair of Majorana states  with  equal
masses  (the mass sign does not matter $\Psi\rightarrow\gamma_5\Psi$
could  be redefined) are equivalent to one four-component Dirac state,
n  is  a  general number of possible physical  spinors  (see
sections 6 and 7).

Dirac  form (9) admits the existence of the  invariant
mass $\mu_{RL}$:
\begin{equation}
\mu_{RL_a}\,^b=\frac 1n\left<\bar \Psi_R^c(x)\Psi_{Lc}(x)\right>
\delta_a\,^b.
\end{equation}

One,  therefore,  may expect here that spontaneous  breaking
will  try  to  cause the least possible symmetry destruction
and  the  diagonal  matrix (15) will be a  solution  to  the
equations  for  the  symmetry  violation  parameters  ("gap"
equations). In addition, formula (15) is necessary  for  our
problem  as  it presents one of the conditions  (Section  5)
that  make  the  existence of "gap" equations possible.  The
remainder  of this section is devoted to the explanation  of
this statement.

Properties  (11)  and  (12) result  in  the  following
equation for the product of the operators $\bar \Psi_R^a\Psi_{Lb}$:
\begin{equation}
\bar \Psi_R^a(x)\Psi_{Lb}(x)=h_{bb'}\bar \Psi_L^{b'}(x)
\Psi_{Ra'}(x)h^{a'a}.
\end{equation}

For  real  vacuum averages of these quantities, the elements
of     $\mu_{{RL}_a}\,^b$ become related:
\begin{equation}
(\mu_{RL})_a\,^b=h_{bb'}(\mu_{RL})_{a'}\,^{b'}h^{a'a}\,,
\end{equation}
where $\mu_{RL}$ is an arbitrary real matrix that is related to the
matrix  $\mu_{LR}$ as follows: $\mu_{RL_a}\,^b\equiv \mu_{LR_b}\,^a$
. For h from  (14),  Eq.
(17)  is  automatically satisfied by the diagonal form  (16)
but  leads  to  $n^2/2$  ratios  for  arbitrary  forms  of
$\mu_{RL}$ matrices.

An   additional  $n^2/4$  relations  for  each  of   the
symmetrical matrices $M_{RR}$ and $M_{LL}$ result from Eq.(13). One
     would argue that this may prevent appearance of a symmetrical $(2n
\times 2n)$ matrix with the properties under discussion from  any
system  of  "gap"  equations in the problem  of  spontaneous
breaking:
\begin{equation}
M=\left|\begin{array}{ll}
M_{RR}& \mu_{RL}\\
\mu_{LR}& M_{LL}\end{array}\right|
\end{equation}
Indeed, "gap" equations should be formulated for each of the
matrix $M$ elements. Matrices $M_{RR}$ and $M_{LL}$ are general matrices
with  non-diagonal elements. The system, then,  consists  of
$n(2n+1)$ equations for the matrix $M$ elements, supplemented
with  $n^2$  relations (13), (17). The number of  variables  in
this system is $n(2n+1)$:  $n(2n-1)$ free parameters of  the
orthogonal matrix diagonalizing $M$ and $2n$ of its eigenvalues.
The  number  of  variables  is  less  than  the  number   of
equations.

In  the  critical  problem,  a  solution  exists  upon
reaching   a  certain  "critical"  value  of  the  effective
interaction  force  characteristics (such  as  the  coupling
constants  in  the Nambu-Jona-Lasinio model,  NJL  \cite{7});  at
that,   the   critical  parameters  should   be   determined
unambiguously.  The number of governing  equations  must  be
equal to the number of parameters. One therefore has to look
for a way to change the ratio between these two numbers.

This can be achieved if  there exists some symmetry of
interactions forming "gap" equations, in which case some  of
$\Psi$,  as well as $V\Psi$, appear to be solutions to the system.  Of
interest  for a real set of equations are real groups  only.
For the critical parameters to be defined unambiguously, the
overall  number  of equations (including those  noninvariant
with  respect  to  $V$)  may  exceed  the  number  of  unknown
quantities  by  the  number of free  parameters  $V$.  Another
option  is: The set of basic equations results in  solutions
for  which additional relations are fulfilled automatically;
this, for example, happens if part of corresponding elements
vanishes.  A  similar situation takes place if the  diagonal
form (15) is the only possible solution for the $\mu_{RL}$ part  of
the  whole matrix $M$. At the same time, the matrices $M_{RR}$  and
$M_{LL}$  can take neither the invariant form of Majorana  masses
nor  the diagonal form (see the last portion of Section  4),
and  "gap" equations need to be formulated for each  of  the
matrix $M$ elements.

In  all  cases when the set of equations  may  have  a
solution,  one  should expect that the  dynamics  themselves
will bring the system to required values, because the region
under  consideration will correspond to the energy  minimum.
In   the  sections  that  follow,  we  will  see  that   the
requirement of solution unambiguity imposes rigid conditions
on  the  choice of the system with the symmetry  $Sp(n/2)$  in
which the proposed mechanism is able to work.

\section{ Gauge-Invariant Model for  Majorana Flavours}

      The  local  gauge  interaction with the  vector  meson
$F_\mu^A(x)$ seems to be the most preferable way for incorporating
$Sp(n/2)$  into  the  problem of Majorana masses.  For  purely
vector  interactions  of massless fermions,  in  particular,
mass  creation  can  only  be associated  with  the  dynamic
violation of chiral (and flavour) symmetry, i.e.,  with  the
generation of vacuum averages (10) and (15).

Currents that define the vector interaction $F_\mu^A(x)$  with
chiral fermions $\psi_{(R,L)}^a(x)$:
\begin{equation}
j_{(R,L)\mu}^A(x)=\bar \psi_{(R,L)}^a(x)\gamma_\mu t_a^{Ab}\psi_{(R,L)b}
(x)
\end{equation}
are directly rewritten with the Majorana operators $\Psi_{(R,L)}(x)$
(4).  From  formulae (4), (2), and anticommutation $\psi_{(R,L)}(x)$
we obtain:
\begin{equation}
J_{(R,L)\mu}^A(x)=\bar \Psi_{(R,L)}^a(x)\gamma_\mu t_a^{Ab}\Psi_{(R,L)b}
(x)\equiv 2j_{(R,L)\mu}^A(x).
\end{equation}
      Note that the axial current of the Majorana states (4)
with matrices having property (2) identically vanishes:
\begin{equation}
J_\mu^{(5)A}(x)=\bar \Psi(x)\gamma_\mu\gamma_5t^A\Psi(x)=0
\end{equation}
for the $R-$ and $L-$systems.

For currents with matrices symmetrical with respect to
(3)  (unity matrix or matrix  $\sigma^P$, see Appendix 2), we observe
the opposite situation: vector currents formed with Majorana
operators  are  equal to zero, whereas  axial  currents  are
equal to chiral vector currents.

Direct  solution  of the dynamic spontaneous  breaking
problem  is  not  attainable under such a system.  For  less
complex  fermion  models,  solving  this  problem  has  been
attempted numerous times both analytically \cite{8} and by  means
of lattice computations \cite{9}.

We  are  interested  in  the  symmetry  properties  of
interactions between Majorana particles. Dependent on  these
properties  is the unambiguity of the solution of  equations
for spontaneous violation parameters, i.e., "gap" equations.
Two  mechanisms are simultaneously engaged in  the  problem:
vector  particle  $F_\mu^A$  mass  generation  and  fermion   mass
generation.  In  NJL models \cite{7}, these two mechanisms,  most
likely interrelated, are represented by various combinations
of Feinman diagrams.

The  effective potential between Majorana fermions  of
interest  to  us  would be easy to determine if  integration
over  field  $F_\mu^A(x)$  could be fulfilled  in  the  functional
integral  for amplitudes. For $R-$ and $L-$operators used  as  $\Psi$
and $\bar \Psi$,  the  solution will depend on combinations (local  and
unlocal) of the following type:
$$\bar \Psi^aZ_1\Psi_a\,, \bar \Psi^aZ_2\Psi_b\bar \Psi^b
Z_3\Psi_a\,,\ldots$$
Operators  $Z$  do  not  contain  indices  $Sp(n/2)$.  At  that,
quantities such as $h^{ab}\Psi_a^T\ldots\Psi_b$  will be transformed
to $\bar \Psi^a\ldots\Psi_a$  using  Majorana formulae (11,
12), and products $t_a^A\,^bt_c^A\,^d$ will bring  us  to the
same result if we use
the formulae given below (23)  or  in Appendix  2  (A.8 and A.9).
This circumstance is responsible  for  the  difference  between
investigations  using chiral and Majorana operators.  Up  to this
point,  the  two notations were in a  form  of  simple variable change.
In Majorana terms, additional real symmetry is  achieved which can make
possible the solution for "gap" equations of spontaneous breaking
(Section 5).

These  equations are, therefore, formed by interaction
with  the real, globally invariant group $O_L(n) = O_R(n)\equiv O(n)$
      with $n(n-1)/2$ arbitrary parameters.

To  make  it more clear,
      let us consider $V_{eff}$ in  the second  order,  with  the
      mechanism of mass generation $F_\mu^A$ isolated, by introducing
      an auxiliary scalar field with non- zero vacuum averages, as
described in Appendix 1.  If  the mass $M_F$ is heavy, the effective
interaction at energies much lower than $M_F$ will be the coupling
"current $\times$ current"(for $R\times R$ and $L\times L$):
\begin{equation}
V_{\rm eff}=-\frac{g_F^2}{2M_F^2}j_\mu^A(x)j^{A\mu}(x)=-\frac{g_F^2}
{8M_F^2}J_\mu^A(x)J^{A\mu}(x).
\end{equation}
This  formula  can be identically transformed  by  means  of
relations  (such  as  the Firz relations)  for  $Sp(n/2)$.  We
obtain:
\begin{equation}
\sum_{A}t_a^{Ab}t_c^{Ad}=\frac 14 \left(\delta_a^d\delta_c^b-h_{ca}
h^{bd}\right).
\end{equation}
Eq. (23) is worked out in Appendix 2 (Formula A8).

      Using  (23)  and the Firz identities for  the
$\gamma\times\gamma$  product  between spinors $RR$, $LL$, and $RL$, we
obtain (omitting arguments x in the operators):
\begin{equation}
V_{RR}=\frac{g_F^2}{4M_F^2}\left(\bar \Psi_R^a\frac{1-\gamma_5}{2}
\Psi_{Rb}\right)\left(\bar \Psi_R^b
\frac{1+\gamma_5}{2}\Psi_{Ra}\right)\,=
\end{equation}
$$=\,\frac{g_F^2}{16M_F^2}\left[\left(\bar \Psi_R^a\Psi_Rb\right)
\left(\bar \Psi_R^b\Psi_{Ra}\right)
-\left(\bar \Psi_R^a\gamma_5\Psi_{Rb}\right)
\left(\bar \Psi_R^b\gamma_5\Psi_{Ra}\right)\right]$$
A  similar  result is obtained for $V_{LL}:\Psi_R
\rightarrow\Psi_L$. Transforming
(22)  into  (24) takes into account equations (7) and  (21).
For the product of currents $R\times L$, we have:
\begin{equation}
V_{RL}=-2\frac{g_F^2}{4M_F^2}\left[\left(\bar \Psi_R^a\frac{1-\gamma_5}
{2}\Psi_{La}\right)\left(\bar \Psi_L^b\frac{1+\gamma_5}{2}\Psi_{Rb}
\right)\right.\,-
\end{equation}
$$-\,\left.\left(\bar \Psi_R^a\frac{1-\gamma_5}{2}\Psi_{Lb}\right)
\left(\bar \Psi_R^b\frac{1+\gamma_5}{2}\Psi_{La}\right)\right].$$
Majorana  conditions (11,12) permit us to prove symmetry  of
Eq.(25) relative to the transposition $R\leftrightarrow L$.

In a single-flavour system, the difference of the signs
in  formulae (24) and (25) would have an important  physical
meaning:  repulsion in the particle-particle  (antiparticle-
antiparticle)  and  attraction in the  particle-antiparticle
system.  In  the  multi-flavour  nonabelian  problem,   this
meaning  is lost; however, it provides evidence against  the
appearance of diagonal $RR-$ and $LL-$ mass matrices in the system
under consideration (Abelian version).

The above physical argument also indicates that in the
gauge  flavour symmetry scheme there probably is not such  a
variant  of  the  neutrino  spectrum   where  neutrino  mass
differences  are  much  less than  masses  themselves.  That
situation  may  take  place when  all  three  roots  of  the
eigenvalue  equation  for  the  Majorana  mass  matrix   are
approximately equal (see Section 7), which is only  possible
when this matrix has a near-diagonal form (Eq. (43)).

\section{Reconciliation of conditions for Matrix $M$ Spontaneous
Appearance}

     The set of "gap" equations uses independent elements of
the orthogonal matrix diagonalizing (18), and eigenvalues $M$,
i.e.,  Majorana  masses, as sought critical parameters.  Our
real   mass   matrix  problem  is  limited  to   only   real
transformations.

All  factors  used, including particle propagators  in
Feynman graphs (if spontaneous violation equations are built
similarly  to  NJL models \cite{7}), are expressed through  these
unknown  quantities.  Overall, there are $n(2n + 1)$ equations
and $n(2n + 1)$ parameters.

On the other hand, there also are relations (13), (17)
imposed on the matrix elements by the "Majorana-ness" of $\Psi_R$,
and $\Psi_L$. These "spare" equations need to be compensated  for
by  tuning conditions as described at the end of Section  3.
There are a few options available here.

      Firstly,  interactions between Majorana  fermions  are
invariant with respect to orthogonal transformations $O(n)  =
O_R(n)  =  O_L(n)$, so that the total number of  equations  may
exceed the number of variables by $n(n-1)/2$, i.e., the number
of the independent elements of the $O(n)$ group.

      Secondly,  spontaneous breaking creates  an  invariant
form  of  the matrix $\mu_{RL}$, or the diagonal matrix  (15).  The
system  will  try  to  cause  the  least  possible  symmetry
breaking.  Under  Majorana-ness conditions (11),  (12),  the
matrix  is symmetric: $\mu_{RL} = \mu_{LR}$. In this case, therefore,  a
part  of $n^2/2$ conditions (17) repeat the part of $n(n+1)/2$
equations  for   elements of symmetrical matrix  with  equal
diagonal terms. It is these $n(n+1)/2$ equations that should
be included in the whole set.

Thirdly,  there  are  two  variants  that  should  be
considered for $M_{RR}$ and $M_{LL}$:

a)    $R,L-$symmetry. Spontaneous creation of both  Majorana
     mass matrices $M_{RR}$ and $M_{LL}$, with the matrices being
identical and $R,L-$symmetry, hence parity, maintained.

b) $R,L-$asymmetry. Let us assume that $M_{LL} = 0$
      (or $M_{RR}  = 0)2$
\footnote{We choose $M_{LL}=0$. Such a choice corresponds to usual
consideration of the see-saw mechanism and seems to facilitate weak
interaction insertion.}.

     This condition is not a result of solution. Therefore, the number
     of equations for $M_{LL}$ that should be considered in this
     variant is $n(n+1)$: $n(n + 1)/2$ conditions for  the elements of
     the symmetric matrix $M_{LL_a}^b = 0$ and $n(n + 1)/2$ "gap"
     equations for these elements from the general system for $M$  (at
     that, $n^2/4$ additional conditions  (13)  are automatically
     fulfilled).  Spontaneous violation of both $R,L-$ symmetry and
     parity takes place.

Let  us write  out  the resultant
     relationships:  the number  of equations minus the number of
      variables is  equal to the  number  of independent symmetry
parameters  of  the equations. We have:
\begin{equation}
\frac{n(n+1)}{2}+\frac{n^2}{4}+\left\{\begin{array}{c}
\frac{n(n+1)}{2}+\frac{n^2}{4}\\
n(n+1)\end{array}\right\}
+\frac{n(n+1)}{2}-n(2n+1)=\frac{n(n-1)}{2}
\end{equation}
for  both the (a) and (b) variants. For the symmetry variant
(a), where $M_{RR} = M_{LL}$, we obtain $(n\neq 0)$:
\begin{equation}
                    n = 2\,.
\end{equation}
For asymmetry variant (b), where $M_{RR}\neq=0$, $M_{LL}=0$, the
solution is
\begin{equation}
n   =   6\,.
\end{equation}
Other  variants  of  choosing conditions  do  not  give  any
physically plausible results.

      It is worth being mentioned again (see Section 3) that
if  the  "gap"  equations do have a solution, this  solution
should  describe a certain energy minimum, and the  dynamics
themselves  will guide the system into the right  region  of
parameter space.

The  solution  $n=6$  , Eq.(28),  is  of  particular
interest. The equation $M_{LL} = 0$ is a necessary condition  for
the  see-saw mechanism to step in (see Section 6). We should
also  remember  that the Majorana matrix  $M_{RR}$  has  pairwise
similar  (in module) eigenvalues. Then, if the  scale  of $M$
masses for $M_{RR}$ is much higher than the Dirac $\mu$ , $n=6$ means
that the scheme contains three Dirac neutrinos of tiny mass,
$\sim \mu^2/M$, and three of huge mass, $\sim M$. We will clarify these
statements in Section 6, while concluding this section  with
the following remark.

Destruction of the $O(n)$ group (as a result  of  fixing
its parameters while solving the equations) will not lead to
the  generation of new Goldstone states. The $O(n)$  group  is
the  symmetry group of the low-energy $V_{eff}$, and within  the
framework of the whole problem it represents a part  of  the
fully  broken $Sp(n/2)$ group. Consequently, Goldstone  states
$O(n)$  must  have already been absorbed by the  formation  of
heavy  masses  for  $F_\mu^A$.  In  the variant  (b)  $R,L-$symmetry
violation, a massless Goldstone state is also absent.  Here,
similarly  to the $U(1)-$problem of QCD cite{10}, one  observes  a
neutral   chiral  anomaly  with  $F_\mu^A-$particles.  A  possible
Goldstone particle will be massive.

\section{  Diagonalization of the Matrix $M$ and Transition  to
Dirac States}

      Let  us consider the variant $n = 6$, $M_{LL} = 0$, the  most
interesting from the physical point of view. We assume  that
the  Majorana mass scale is much larger than the Dirac  mass
scale:
\begin{equation}
M\gg \mu\,.
\end{equation}
In this case, the matrix $M$, Eq. (18), is easy to diagonalize
in two steps. At first, we diagonalize $M_{RR}$ by the orthogonal
matrix $U$, i.e., with the transformation:
\begin{equation}
\Psi'_R=U\Psi_R\,.
\end{equation}
Simultaneously, a transition to $\Psi'_L = U \Psi_L$ can be  made,  in
which  case neither the diagonal $\mu_{RL}$ nor $M_{LL} = 0$ appears  to
change. The $(12\times  12)$ matrix that results is:
\begin{equation}
U^TMU=\left|\begin{array}{cccccc}
M_{R1}&      & 0   &\mu&& 0\\
      &\ddots&     &   &\ddots&\\
0 && M_{R6}& 0 && \mu\\
\mu && 0 & 0& & 0\\
      &\ddots&     &   &\ddots&\\
0 && \mu & 0 && 0\end{array}\right|\,.
\end{equation}
This matrix splits into a product of twofold matrices we are
well  familiar with from works on the see-saw mechanism (see
review \cite{3}):
\begin{equation}
m^D=\left|\begin{array}{cc}
M_D & \mu\\
\mu & 0\end{array}\right|,\qquad D=1, 2, \ldots, 6.
\end{equation}

      At  $\mu\ll  M_D$, two eigenvalues of $m^D$ differ from  each
other in magnitude by a factor of $m^2/M_D^2$:
\begin{equation}
\lambda_{1,2}^D=\frac{M_D}{2}\pm\sqrt{\frac{M_D^2}{4}+\mu^2}\simeq
\left\{\begin{array}{c}
M_D+\frac{\mu^2}{M_D}\\
-\frac{\mu^2}{M_D}\end{array}\right..
\end{equation}
Formula  (33) is valid for any sign of $M_D$.  Note  that  this
sharp distinction $\lambda$ may result only if the second element on
the  principal diagonal in (32) is equal to zero,  i.e.,  at
the corresponding mass $M_L^{(D)} = 0$. This is the physical meaning
of  the  condition $M_{LL} = 0$ in Section 5. For variant (a)  in
Section  5,  $M_{LL}\neq 0$ and even $M_R = M_L$, the spectrum  of  two
neutrinos $(n = 2)$ will  have an absolutely different form.

Eigenfunctions (32) are determined by rotation of  the
matrix orths through the transformation:
\begin{equation}
V_D=\left|\begin{array}{cc}
\alpha_D &\beta_D\\
-\beta_D &\alpha_D\end{array}\right|,
\end{equation}
where quantities $\alpha_D$ and $\beta_D$ are equal to:
\begin{equation}
\alpha_D=\frac{1}{\sqrt{1+(\mu/M_D)^2}}\,,\qquad
\beta_D=\frac{\mu/M_D}{\sqrt{1+(\mu/M_D)^2}}\,,\qquad
\alpha^2+\beta^2=1\,.
\end{equation}
The signs in Eq. (35) are chosen for further convenience.

     In matrix (31), let us place masses $M_{Ri}$ so that $M_{R6} = -
M_{R1}$,  $M_{R5} = - M_{R2}$, $M_{R4} = - M_{R3}$. Then, for mass
matrix (18), we  have  three pairs of heavy masses: $M_{\pm D}$, $D =
1,2,3$, and three pairs of light masses, $m_{\pm D}$, ($\pm D$ means
the sign of the mass $D$). The eigenfunctions of diagonalized states
are:
\begin{equation}
\Psi_{\pm D}=U_{\pm D}\,^a\left(\alpha_{\pm D}\Psi_{Ra}+\beta_{\pm
D}\Psi_{La}\right),
\end{equation}
$$
\psi_{\pm D}=U_{\pm D}\,^a\left(-\beta_{\pm D}\Psi_{Ra}+\alpha_{\pm
D}\Psi_{La}\right).$$
Depending on the sign selected (35), we have:
\begin{equation}
\alpha_D=\alpha_{-D}\,,\qquad\quad \beta_D=-\beta_{-D}\,.
\end{equation}

      Wavefunctions (36) have a property similar to Majorana
relations  (11),  (12).  Using  these  relations,  one   can
establish  a  connection between states (36) with  different
mass signs:
\begin{equation}
\gamma_5 hC\bar \Psi^{TD}=\Psi_{-D}\,,
\end{equation}
$$\gamma_5 hC\bar \psi^{TD}=-\psi_{-D}\,.$$
Mass   disposition  is  chosen  so  that  $h_{D'D}\equiv h_{-DD}$.   We,
therefore,  can  continue using covariant formulae,  raising
and lowering the diagonal indices $D$. In order to prove (38),
we first write:
\begin{equation}
\gamma_5hC\bar \Psi^{TD}=hU^{+T}h^+\left(\alpha_D\gamma_5hC\bar
\Psi_R^T+\beta_D\gamma_5hC\bar \Psi_L^T\right).
\end{equation}
From  formula (13) written in the covariant form, it follows
that there is a relation between the elements of matrices  $U$
that diagonalize $M_{RR}$:
\begin{equation}
h_{D'D}U^{+TD}\,_bh^{ba}=\pm U_{-D}\,^a\,;\quad D'=-D\,;\quad
a, b=1,2,\ldots,6\,.
\end{equation}
Taking into consideration (11), (12) and sign change  for $\beta$
with  the  sign  change for the mass $M_D$, we  finally  obtain
(38).

Equations  (38)  facilitate construction  of  physical
Dirac  states  with positive masses. Let us first  construct
true  Majorana  states $(\psi=C\bar \psi^T)$ for each  of  the  diagonal
states $D = 1,2,3$. We have
\begin{equation}
\chi_1^{(M_D)}=\frac{\Psi_D+C\bar \Psi^{TD}}{\sqrt{2}}=
\frac{\Psi_D+\gamma_5h^+\Psi_{-D}}{\sqrt{2}}\,,
\end{equation}
$$\chi_2^{(M_D)}=\frac{\Psi_D-C\bar \Psi^{TD}}{\sqrt{2}i}=
\frac{\Psi_D-\gamma_5h^+\Psi_{-D}}{\sqrt{2}i}\,,$$
which  will  be  similar also for states with  light  masses
$\chi_1^{(m_D)}$ and $\chi_2^{(m_D)}$.

Massive Dirac states with positive masses are expressed
as follows:
\begin{equation}
\Psi=\frac{\chi_1+i\chi_2}{\sqrt{2}},\qquad
\bar \Psi=\frac{\bar \chi_1^T-i\bar \chi_2^T}{\sqrt{2}}C^{+T},
\end{equation}
for any $M_D > 0$, $m_D > 0$.

It  is  well known that transformations (41) and  (42)
transfer  the  free Lagrangians for Majorana particles  into
the  Lagrangian  for Dirac particles. Thus, twelve  Majorana
states will contribute to three heavy and three light  Dirac
particles.

\section{ Mass Matrix $M$ Spectrum and Light Neutrino
Masses}

      Let  us  evaluate what kind of spectra can be obtained
from matrix (18) in the scheme under consideration at $n = 6$,
$M_{LL}  = 0$, and the diagonal Dirac form $\mu_{RL}$. The Majorana mass
matrix  $M_{RR}$ obeys conditions (13). Let us assume that scales
$\mu\ll  M$,  in order to have the states split into light  and
heavy  masses  (the see-saw mechanism).  Then,  $M_{RR}$  can  be
written as:
\begin{equation}
M_{RR}=M\left|\begin{array}{rrrrrr}
a_{11}& a_{12}& a_{13}& a_{14}& a_{15}& a_{16}\\
a_{12}& a_{22}& a_{23}& a_{24}& a_{25}& -a_{15}\\
a_{13}& a_{23}& a_{33}& a_{34}& -a_{24}& a_{14}\\
a_{14}& a_{24}& a_{34}& -a_{33}& a_{23}& -a_{13}\\
a_{15}& a_{25}& -a_{24}& a_{23}& -a_{22}& a_{12}\\
a_{16}& -a_{15}& a_{14}& -a_{13}& a_{12}& -a_{11}
\end{array}\right|.
\end{equation}
We have twelve independent elements that take certain values
imposed by the $gap$ equations. Since the equations are  not
known  to us, we have to limit ourselves to a rough estimate
of values resulting from the matrix form (43).

Let us take all independent elements $M_{RR}/M\simeq 1$. There
is  no reason to think that the equations with high symmetry
will  result in parameters being essentially different  from
each other. For these parameters we cannot imagine any other
physical justified distribution pattern .

The only difference inherent in (43) is the difference
in  signs: $\pm 1$  (in $M$ units), so one can take  independent
elements  with  opposite  signs.  We  have  checked  several
variants,  all  of which lead to eigenvalue  equations  with
coefficients alternating in sign. The second coefficient  of
the equation is obviously negative (at $(M_D^2)^2$).

If  all  independent  elements  $a_{ik}=1$,  then  the
eigenvalue equation $M_{RR}$ is written as $(x=M_D^2/M^2)$:
\begin{equation}
x^3-18x^2+48x-32=0\,,
\end{equation}
with  roots $x_1\simeq 1.02$, $x_2=2$, and $x_3\simeq 14.92$. Light
masses squared (in $(\mu^2/ M)^2$ units) are:
$$m_1^2\simeq 0.067,\qquad m_2^2=0.5,\qquad m_3^2\simeq 1.$$
Experimentally obtained ratio \cite{1}, Eq. (1)
\begin{equation}
\left|\frac{\Delta m_{23}^2}{\Delta m_{12}^2}\right|=\frac{m_3^2-
m_2^2}{m_2^2-m_1^2}\simeq 3,
\end{equation}
is, of course, far from the experimental value in Section 1.

      Other  variants of $a_{ik}=\pm 1$ lead to various equations.
The eigenvalue equation
$$x^3-18x+80x-64=0$$
results if the matrix elements are negative: 1) $a_{13} = -1$; 2)
$a_{14}  =  -1$; 3) $a_{13} = -1$; and so on. Ratio (45) here is equal
to $\sim 10$. The equation
$$x^3-18x^2+64x-32=0$$
can  be obtained at 1) $a_{22} = -1$; 2) $a_{22} = -1$, $a_{16} = -1$;
and so on. Ratio (45) here is equal to $\sim 8$.

      One can obtain
      equations with one real root, which  is physically unacceptable:
$$x^3-12x^2+96x-D=0$$
$D=32$ for $a_{14}= a_{15} = a_{13} = -1$; $D = 126$ at $a_{12}=-1$; $D
      = 136$ at $a_{14} = a_{15} = a_{13} = a_{16} = -1$.

      Finally, there is a frequent situation when the  least
root  $x$ appears to be smaller than 1, whereas the other  two
roots  are larger than 1. Then, ratio (45) is big, which  is
required phenomenologically (see \cite{1}). The equation
\begin{equation}
x^3-18x^2+80x-D=0
\end{equation}
is valid for $D = 32$, $a_{12} = a_{15} = -1$, roots $x_1\simeq 0.045$,
 $x_2\simeq 6.51$, $x_3\simeq 11.04$ and ratio (45) equal to $\simeq
35$; for $D = 24$, $a_{23} = a_{15} = -1$, roots $x_1\simeq 0.325$,
$x_2\simeq 6.8$, $x_3\simeq 10.9$, ratio (45) equal to $\simeq 53$. The
coefficient values being high, the small quantity $x_1$ coexists with
the rather large $x_2$ and $x_3$, facilitating fitting big ratios (45).

We note that big numbers (45) are relatively easy  to obtain  in
matrices of the type under consideration.  The character of the
spectrum obtained experimentally, where one neutrino  state  is
located  far from  the other,  almost degenerate   ones,   means,  in
terms   of the present investigation, that $x_1 < 1$, while the other
two $x_{2,3}> 1$.  This  kind of situation is not rare in matrices of
the (43) type and there is nothing unique to it.

The   light  particle
mass spectrum  shows  normal hierarchy:   the  heavy  mass corresponds
to   the   large difference $\Delta m_{23}^2$.  The small difference
$\Delta m_{12}^2$ is achieved with   the   Majorana  masses
$x_{2,3}> 1$.  At that,   no degeneration of states 1,2 occurs; what
is only noted is the effect  of difference for inverse squares of
large  numbers.  In
terms of the Majorana spectrum, the situation is  quite ordinary.

\section{ Conclusion}

       The scheme under consideration, if ever possible  \cite{8,
9},  does not include many aspects of the phenomena that are
essential under a less simplified model (renormalization  of
nonperturbative  solutions,  scale  dependence  of   various
factors).  On the other hand, it has a number of  attractive
features. Of interest is the connection in (25) of the light
neutrino number and spectrum with the compulsory spontaneous
breaking  of  the  $-R,L-$symmetry. One needs  to  gain  deeper
insight into the meaning of this connection.

Spectrum reproduction indicates that the reason for the
specific  neutrino mass disposition is rather  banal:  the
smallness  of one of the Majorana masses (using  comparative
units)  $M_1^2 < 1$ and the relatively heavy two others $M_{2,3}^2>
1$.  This scheme results in normal mass hierarchy of neutrino
states.

   Mixing angles for observed neutrino flavours can not be
considered  unless the mechanism of mass spectrum generation
is determined for charged leptons. The next step, therefore,
is  to  include charged, and, as is well known,  exclusively
Dirac leptons in the investigation of weak interactions.  In
this  connection, the $R,L-$symmetry violation,  which  occurs
spontaneously,  appears  to  be  significant  as  it   means
symmetry violation outside the Standard Model.

This  phenomenon not taken into account, one has  to
overcome  the following difficulty. The Standard  left  weak
current  constructed  by  means of chiral  states  $\psi_L$,  when
written in terms of Majorana particles and then in terms  of
massive Dirac particles, appears to also include their right-
hand components (or antiparticles):
\begin{equation}
\bar \psi_L^a\gamma_\mu\psi_{La}\equiv\frac 12\bar \Psi_L^a\gamma_\mu
\gamma_5\Psi_{La}\,,
\end{equation}
since $\bar \Psi_L^a\gamma_\mu\Psi_{La}\equiv 0$
(see  section  4).  Including  only   left
components  of  physical Dirac states $\psi_D$, $\Psi_D$,  in  the  weak
current, which is attractive from the phenomenological point
of  view, would break the $Sp(n/2)-$symmetry once again.  This
would mean that the joint $(Sp(n/2))$ and weak Standard models
are non-renormalizable.

      The  author  would  like  to  thank  D.~I.~Dyakonov  for
essential help and Y.~I.~Azimov for useful comments.
\newpage

{\bf Appendix 1. Mass of Gauge Boson $F_\mu$}

      In  order the vector boson $F_\mu^A$, $(A = 1,2,\ldots,n(n +
1)/2)$ acquires  a mass, it is sufficient that the scalar field  is
present  under the same adjoint representation as  $F_\mu^A$.  The
scalar  field $\phi^A$  is expressed by means of  the   symmetric
tensor $\phi_{ab}$ (or $\phi^{ab}=h^{aa'}h^{bb'}\phi_{a'b'}$.):
$$
\phi^A=t_a^A\,^bh_{bb'}\phi^{ab'}=\phi_{ab'}h^{b'b}t_b^A\,^a\equiv
t_a^A\,^b\phi_b\,^a\,. \qquad\qquad  (\mbox{A.1})
$$
All   of   these  tensors  have  $N=n(n+1)/2$  independent
components, $n(n+1)$ real parameters.

The  Standard  Higgs  Lagrangian,  which  is  $Sp(n/2)$ -
invariant
$$\phi_a^{+b}\left(\partial_\mu\delta_b^c+it_b^{Ac}F_\mu^A(x)\right)
\left(\partial_\mu\delta_c\,^d-it_c^{A'}\,^dF_\mu^{A'}\right)
\phi_d\,^a(x)-V(|\phi|)\,,\qquad\qquad\qquad\qquad (\mbox{A.2})$$
with  simple  $V(|\phi|)$ chosen so that
$|\phi|^2=\phi_a\,^{+b}\phi_b\,^a$ acquires the vacuum average
$$\left<\phi_a\,^{+b}\phi_b\,^{a'}\right>=\frac 1N\eta^2\delta_a\,^{a'}
,\qquad\qquad\qquad\qquad\qquad\qquad (\mbox{A.3})$$
creates  masses for all $F_\mu^A$. At that, all phases  of $\phi$  are
absorbed in the formation of heavy masses for $F_\mu^A$,  so  that
$\phi^A$  are  present  as really existing particles only at  very
high  energies.  Large number of possible $Sp(n/2)$  invariant
forms permits to use, besides (A.3), also
more complicated  spectra of heavy $\phi$ particles.
\newpage

{\bf Appendix 2.}

        Hermitian matrices $t^A$ in the representation n of the
$Sp(n/2)$  group  can  be selected from $n^2-1$  infinitesimal
operators $T_A$ in the fundamental representation $N=n$ of  the
better known group $SU(N)$. The even-dimensional operators $T_A$
can  be  constructed  to  form two groups  with  respect  to
operation (2) with the skew antisymmetrical matrix $h(n\times
n)$, Eq. (14):
$$h^+t^Ah=-t^{AT}\,,\qquad h^+\sigma^Ph=\sigma^{PT}\,.
\qquad\qquad\qquad\qquad  (\mbox{A}.4)$$
We  have  $n(n + 1)/2$ matrices $t^A$ and
$\left[\frac 12 n(n-1)-1\right]$
matrices $\sigma^P$ \cite{4}.  The  operators $t^A$,
$\sigma^P$ and unity operator I form the complete basis for space of
$(n\times n)$ matrices, with its usual normalization being:
$$Trt^At^B=\frac 12 \delta^{AB}\,,\qquad
Tr\sigma^P\sigma^Q=\frac 12\delta^{PQ}\,,\qquad
Tr t^A\sigma^P=0\,.\qquad\qquad\qquad  (\mbox{A}.5)$$
Expanding the right parts of commutators and anticommutators
$t$ and $\sigma$  in  the complete basis and using Eqs.  (A.4),  we
obtain:
$$[t^A,t^B]=if^{ABC}t^C,\qquad \{t^A,t^B\}=\frac{1}{2n}
\delta^{AB}I+d^{ABP}\sigma^P,$$
$$[t^A,\sigma^P]=if^{APQ}\sigma^Q,\qquad\qquad\qquad\qquad
(\mbox{A}.6)$$ $$[\sigma^P,\sigma^Q]=if^{PQA}t^A,\qquad
\{\sigma^P,\sigma^Q\}=\frac{1}{2n}\sigma^{PQ}I+d^{PQR}\sigma^R.
$$
The  coefficients $d^{PQR}$ in the anticommutator $\sigma$ are equal  to
zero,  as  it possible to show. From Eq.(A.6) it is  evident
that the matrices $t^A$ form a closed algebra, which in turn is
responsible  for the representation $n$ of the group  $Sp(n/2)$.
Matrices $\sigma^P$   are  responsible  for  another   irreducible
representation of this group with dimension
$\left[\frac 12 n(n-1)-1\right]$.

       Let   us   use  the  well-known  equation   for   the
infinitesimal  operators  in the fundamental  representation
$SU(n)$.  In  terms  of  the operators  $t^A$  and $\sigma^P$,  we  have
(summation over indices $A$ and $P$):
$$t_a^A\,^bt_c^A\,^d+\sigma_a^{Pb}\sigma_c^P\,^d=\frac 12
\left(\delta_a\,^d\delta_c\,^b-\frac{1}{n}\delta_a\,^b\delta_c\,^d
\right).\qquad (\mbox{A}.7)$$
After applying Eq.(A.4) to both parts of (A.7), changing the
indices,  adding  and  subtracting (A.7)  and  the  equation
obtained, we have:
$$t_a^A\,^bt_c^{Ad}=\frac 14\left(\delta_a\,^d\delta_c\,^b-
h^{bd}h_{ca}\right)            \qquad (\mbox{A}.8)$$
$$\sigma_a^P\,^b\sigma_c^P\,^d=\frac 14\left(\delta_a\,^d\delta_c\,^b+
h^{bd}h_{ca}-\frac 2n \delta_a\,^b\delta_c\,^d \right).      \qquad
(\mbox{A}.9)$$

With  the usual notations, $(t^{A^T})^a\,_b=t_b^{A_a}$, which is used
in the calculations.

\end{document}